\documentclass[a4paper,11pt]{article}
\usepackage[utf8]{inputenc}
\usepackage{amsfonts}
\usepackage{amssymb}
\usepackage{graphicx}
\usepackage{fancyhdr}
\usepackage{fancyvrb}
\usepackage{fancybox}
\usepackage{float}
\usepackage{geometry}
\usepackage{minitoc}
\usepackage{indentfirst}
\usepackage{multicol}
\usepackage[outerbars]{changebar}
\usepackage{pifont,textcomp}
\usepackage{amsfonts}
\usepackage{amsmath}
\usepackage{amscd}
\usepackage{array}
\usepackage{multirow}
\usepackage{mathrsfs}
\usepackage{amssymb}
\usepackage{amsthm}
\title{ Thermodynamics of ideal gas at Planck scale with strong quantum gravity measurement  }
\author{ Latévi Mohamed Lawson
\space\\\\
 African Institute for Mathematical Sciences (AIMS) Ghana\\
 Summerhill Estates, East Legon Hills, Santoe, Accra\\
 P.O. Box LG DTD 20046, Legon, Accra, Ghana\\\\
latevi@aims.edu.gh and lawmenx@gmail.com }

\begin{document}
\maketitle

\begin{abstract}
More recently in [J. Phys. A: Math. Theor. {\bf 53}, 115303 (2020)], we have introduced a set of noncommutative algebra that describes the space-time at the Planck scale.  The interesting significant result we found is that the generalized uncertainty principle induced a maximal length of quantum gravity which has different physical implications to the one of generalized uncertainty principle with minimal length. The emergence of a maximal length in this theory revealed strong quantum gravitational effects at this scale and predicted the detection of gravity particles with low energies. To make evidence of these predictions, we study the dynamics of a free particle confined in an infinite square well potential in one dimension of this space. Since the effects of quantum gravity are strong in this space, we show that the energy spectrum of this system is weakly proportional to the ordinary one of quantum mechanics free of the theory of gravity. The states of this particle exhibit proprieties similar to the standard coherent states which are consequences of quantum fluctuation at this scale.
Then, with the spectrum of this system at hand,  we analyze the thermodynamic quantities within the canonical and microcanonical ensembles of an ideal gas made up of $N$  indistinguishable particles at the Planck scale. The results show a complete consistency between both statistical descriptions. Furthermore, a comparison with the results obtained in the context of minimal length scenarios and black hole theories indicates that the maximal length in this theory induces logarithmic corrections of deformed parameters which are consequences of a strong quantum gravitational effect.

\end{abstract}

{\bf Keywords:}  Generalized Uncertainty Principle; Minimal and   maximal lengths of  quantum gravity; Thermodynamics of ideal gas; Logarithmic corrections

\section{Introduction}

In the past few years,  the Generalized Uncertainty Principle (GUP)  has emerged as a path of finding a consistent quantum formulation of the theory of gravity  \cite{2,3,4}. This GUP obtained by adding small quadratic corrections to the Heisenberg algebra leads to the existence of minimal uncertainties in position or in momentum \cite{5,6}. All the active candidates to the search of these minimal uncertainties such as string theory \cite{7}, black hole theory \cite{8}, loop quantum gravity \cite{9} and quantum geometry \cite{10}  are mostly restricted to the case where there is a nonzero minimal uncertainty in the position.  Only Doubly Special Relativity (DSR) theories \cite{11,12,13,14} suggest an addition to the minimal length, the existence of a  maximal momentum. Recently,  Perivolaropoulos proposed a consistent algebra that induces for a simultaneous measurement, a maximal length and a minimal momentum \cite{15}. In this approach, the maximal length of quantum gravity is naturally arisen in cosmology due to the presence of particle horizons.

 In this prophetic paper \cite{15}, Perivolaropoulos also predicted the simultaneous existence of maximal and minimal position uncertainties. More recently,  without any formal knowledge of this result, we introduced a  version of position-dependent noncommutative space in two-dimensional (2D) configuration spaces which,  for simultaneous measurement lead to minimal and maximal lengths of quantum gravity \cite{1}. The interesting physical consequence we found is that the existence of maximal length in this theory brings a lot of new features to the Hilbert space representation and agrees with the similar perturbative approaches predicted by DSR  theories  \cite{11,13,14}. In this paper \cite{1}, we predicted that this concept of maximal length could induce strong graviton localizations and could be the approach candidate for the measurement of quantum gravity with low energies. In the present paper to make evidence of this prediction, we investigate in 1D,  the outgoing of this GUP model with maximal length in the statistics of the canonical and microcanonical ensembles of an ideal gas at the Planck scale. Since the quantum gravity is strongly measured at this scale,  the thermodynamic quantities induce for both descriptions logarithm corrections of the deformed parameter $\tau$ $(\ln\tau)$. 
 This situation perfectly fits with the obtained results at the extremal limit of a black hole geometry \cite{16,17,18,19}. Comparing these consequences with those of minimal length measurement \cite{20,21,22,23,24,25}, show that the effects of both measurements are fundamentally different. In fact, the minimal length formalism shifts quadratically the thermodynamic quantities at the order of the deformed parameter $\beta_{GUP}$. Thus, at the extremal limit $(\beta_{GUP} \rightarrow 0)$, one recovers the ordinary quantities at this scale while in this framework by tending our deformed parameter $\tau$ to zero, these thermodynamic quantities diverge. We come out with these observations that, the minimal length formalism induces weak quantum gravities while ours induce strong quantum gravities at the Planck scale.

In the present paper, before analyzing the behavior of this ideal gas, which is a classical and a well-known topic, we study the dynamics of a free particle confined an infinite square well potential at the frontier of the Planck scale. We show that the spectrum of this system is weakly proportional to the ordinary one of quantum mechanics without gravity perturbations. This indicates contractions of the energy levels, allowing particles
to jump from one state to another with low energies. These deformations observed at this scale are in perfect analogy with the theory of General Relativity (GR) where the gravitational field becomes stronger for heavy systems that contract the space,
allowing light systems to fall down with low energies  \cite{25a'}.
Since, the quantum gravitational fluctuation becomes important at this scale, the states of this particle exhibit property similar to the Gaussian states of the standard quantum mechanics.   
In the next section, we review in 1D, the GUP with the maximal length and its deformed translation symmetry \cite{1}. Section $\ref{sec3}$, is devoted to the study of a non-relativistic quantum particle in an infinite square well potential at the Planck scale. We give the spectrum of this system by solving analytically the Schrödinger equation. In section $\ref{sec4}$, we deduce from this spectrum a statistical description of an ideal gas in canonical and
microcanonical ensembles. We show that the canonical ensembles based on the use of the partition function and the microcanonical ensemble based on the density of states lead to the same results. The conclusion is given in section $\ref{sec5}$.

\section{ GUP with maximal length and its  deformed translation  symmetry}
\subsection{GUP  with maximal length}
The quantum gravity according to Kempf \textit{et Al} formalism is manifested by the quadratic  deformation of the Heisenberg algebra \cite{2}. Recently, we proposed a new version of position-dependent deformed Heisenberg algebra in two-dimensional configuration spaces that introduces a simultaneous presence of maximal and minimal
position uncertainties \cite{1}. We define this algebra  as follows:

{\bf Definition 2.1:} {\it Given a set of symmetric operators $\hat X, \hat Y, \hat P_x, \hat P_y$  defined on  the 2D Hilbert space   $\mathcal{H}_{\theta\tau}= \mathcal{L}^2(\mathbb{R}^2)$  and satisfy the following commutation relations 
	%and all possible permutations of the Jacobi identities 
\begin{eqnarray} \label{al1}
	[\hat X,\hat Y]&=&i\theta (1-\tau\hat Y +\tau^2 \hat Y^2),\quad [\hat X,\hat P_x ]=i\hbar (1-\tau \hat Y +\tau^2 \hat Y^2),\cr
	{[\hat Y,\hat P_y ]}&=&i\hbar (1-\tau \hat Y +\tau^2 \hat Y^2),\quad\quad
	{ [\hat P_x,\hat P_y]}=0, \cr 
	{[\hat Y ,\hat P_x]}&=&0,\quad [\hat X,\hat P_y]=i\hbar\tau(2\tau \hat Y\hat X-\hat X)+
	i\theta\tau(2\tau \hat Y\hat P_y-\hat P_y),
\end{eqnarray}
where $\theta,\tau \in \mathbb{R}_+^*$  are both deformed parameters and manifest themselves as lengths describing the space at a short distance.}\\
The parameter $\tau$ is the GUP  deformed parameter \cite{3,4,4'} related to quantum gravitational effects at this scale.
The parameter $\theta$ is related to the  noncommutativity of the space at this scale  \cite{5,6,25'}. In the framework of noncommutative classical or quantum mechanics, this parameter  is proportional to the inverse of  constant magnetic   field such as  $\theta=1/B$ \cite{25'a,25'b,25'c}. Since the algebra (\ref{al1}) describes the space  at the  Planck scale, then such  magnefic  fields   are  necessarily  superstrong and  may play the  role of primordial magnetic fields in  cosmological  dynamics  \cite{25''a}. Thus, in this paper, we   unified  both parameters as the minimal length scale $\Delta X_{min}=l_{min}=\tau\theta=\tau/B$. Futhermore, at the same scale, this minimal length measure   $\Delta X_{min}= \tau\theta$ is  coupled with the maximal one  $\Delta Y_{max}=\tau^ {-1}=l_{max}$ by the  inverse of strong magnetic fields $B$ as follows
\begin{eqnarray}\label{2}
\Delta X\Delta Y=\frac{1}{B}= l_{min}l_{max}.
\end{eqnarray}
At $n$-dimensional sets of the algebra (\ref{al1}),  the equation (\ref{2})  indicates a sort of discreteness of the space at this
scale where one has an alternation of minimal lengths
$l_{min}$ with the maximal lengths $l_{max}$. This situation can be compared to a  lattice system in which,
each site represented by $l_{min}$ is spaced by $l_{max}$.
At each singular point $l_{min}$, result of the unification of magnetic fields and quantum gravitational fields. Since the Planck scale marks the frontier of our universe, then the existence of magnetic fields at this scale can only be from a multiverse that bounces at these minimal positions $l_{min}$.  As we have predicted in our previous work \cite{1}, this scenario lets break up the big bang singularity. 
This present part of the paper is currently under investigation and will be the object of further exhaustive study.

 Now, let us start with the particular case.   In one  dimensional  set, the algebra (\ref{al1}) simplifies greatly.
 
 {\bf Proposition 2.1:} {\it Let  Hilbert $\mathcal{H}_{\tau}= \mathcal{L}^2(\mathbb{R})$  be one dimensional  space that describes the 
 	noncommutative space. The symmetric operators $\hat X$ and $\hat P$ that act on this space satisfy the following
 	relation
\begin{eqnarray} \label{alg1}
{[\hat X,\hat P ]}=i\hbar (\mathbb{I}-\tau \hat X +\tau^2 \hat X^2).
\end{eqnarray}
By  setting these operators as  follows 
\begin{eqnarray}\label{reR}
\hat X=\hat x,\quad 
\hat P=(\mathbb{I}-\tau \hat x+\tau^2\hat x^2)\hat p,
\end{eqnarray}
we recover the relation (\ref{alg1}), where the  symmetric operators $\hat x$ and $\hat p$  satisfy the  ordinary Heisenberg algebra 
$[\hat x,\hat p]=i\hbar$.}

 From the representation (\ref{reR}), one can interpret $\hat x$  and $\hat p $ as the set of operators at low energies which
 has the standard representation in position space and $\hat X,\hat P$ as the set of operators at high energies, where they have the generalized representation in position space. Furthermore, the proposal (\ref{alg1}) is consistent with the recent approach of quantum gravity measurements introduced by Perivolaropoulos \cite{15}. In this approach the quantum gravity is naturally arisen from cosmology due to the presence of particle horizons with the parameter $\alpha=\alpha_0\frac{H_0^2}{c^2}$ where  $ H_0$ is the Hubble constant, $c$ is the speed of light and $\alpha_0$ is a dimensionless parameter.	
		From  this commutation relation  (\ref{alg1}), 
		an interesting feature  can be observed through the  following  uncertainty relation:
		\begin{eqnarray}\label{uncertitude}
		\Delta  X\Delta P\geq \frac{\hbar}{2}\left(1-\tau\langle \hat X\rangle+\tau^2\langle \hat X^2\rangle\right).\label{in2}
		\end{eqnarray}
		Using the relation 
		$ \langle \hat X^2\rangle=\Delta  X^2+\langle \hat X\rangle^2$, the equation (\ref{uncertitude}) can be rewritten as a second order equation for $\Delta P$. The solutions for $\Delta X$ are as follows 
		\begin{equation}
		\Delta X=\frac{\Delta P}{\hbar \tau^2}\pm \sqrt{\left(\frac{\Delta P}{\hbar \tau^2}\right)^2
			-\frac{\langle \hat X\rangle}{\tau}\left(\tau\langle \hat X\rangle-1\right)
			-\frac{1}{\tau^2}}.
		\end{equation}
		The reality of solutions gives the following minimum
		value for $\Delta P$
		\begin{eqnarray}
		\Delta P=\hbar \tau\sqrt{1-\tau \langle \hat X\rangle+\tau^2\langle \hat X\rangle^2}.
		\end{eqnarray}
		Therefore, these equations lead to the absolute minimal uncertainty $\Delta P_{min}$ in $P$-direction  and the absolute maximal uncertainty  $\Delta X_{max}$ in $X$-direction  
		for $\langle  \hat X\rangle=0$, such as:
		\begin{eqnarray} 
		\Delta P_{min}=\hbar\tau, \quad 
		\Delta X_{max}= l_{max}=\frac{1}{\tau}\label{max}.
		\end{eqnarray}

It is well-known \cite{2} that, the   existence of minimal uncertainty   raised
the question of singularity of the space i.e space is inevitably bounded by minimal quantity beyond which any further localization of particle is not possible. In the  presence situation, the minimal  momentum $\Delta P_{min}$   leads to the lost of representation in $P$-direction and a maximal measurement $\Delta X_{max}$  conversely will be the physical  space of  wavefunction representations i.e  all functions $\psi (X) \in  \mathcal{H}_{\tau}= \mathcal{L}^2(-\infty,+\infty)$   vanish  at the  boundary $ \psi(-\infty)=0=\psi (+\infty)$. Since particles in $P$-direction  cannot be localized in a precise way. This implies a certain fuzziness of momentum in this direction. The consequence of this fuzziness can be understood by the following corollary.\\

{\bf Corollary 2.1:}
{\it From the representation (\ref{reR}) follows immediately that  the position operator  $\hat X$ is symmetric while the momentum operator $\hat P$  is not
	\begin{eqnarray}
	\quad \hat X^\dag= \hat X, \quad  \hat P^\dag= \hat P+i\hbar\tau(1-2\tau \hat X).
	\end{eqnarray}}
{\bf Proof:} Since the operators $ \hat x,\hat p $ are symmetric then  $\hat X^\dag=\hat x^\dag=\hat X$ and  $\hat P^\dag=\hat p^\dag (\mathbb{I}-\tau \hat x+\tau^2\hat x^2)^\dag=\hat p (\mathbb{I}-\tau \hat x+\tau^2\hat x^2)=(\mathbb{I}-\tau \hat x+\tau^2\hat x^2)\hat p + i\hbar\tau(1-2\tau \hat x)=\hat P+i\hbar\tau(1-2\tau \hat X),$ finishing the proof.\\
	
	In  order to guarantee the  symmetry of this operator, we arbitrary restrict the  study
	from the infinite-dimensional Hilbert space $ \mathcal{H}_{\tau}$ into its  bounded dense Domaine $ \mathcal{D}_\tau= \mathcal{L}^2(-l_{max},l_{max})$ in such a way that,  for $\tau\rightarrow 0$ one recovers the entire space  $ \mathcal{H}_{\tau}$.  This restriction perfectly fits with the work of Nozari and Etemadi done in momentum space \cite{25''b}.  In addition of this condition,  we propose the following deformed completeness relation to get the symmetry of the operator $\hat P$.\\		

  {\bf Proposition 2.2:} {\it For the given  complete basis $\{|x\rangle\}\in \mathcal{D}_\tau$  such as
\begin{eqnarray}
\int_{-l_{max}}^{+l_{max}}\frac{dx}{1-\tau x+\tau^2 x^2}|x\rangle \langle x|=\mathbb{I}.
\end{eqnarray}
we have 
\begin{eqnarray}
 \langle \phi|\hat P\psi\rangle = \langle \hat P^\dag \phi|\psi\rangle,
\end{eqnarray}
 such as 
 \begin{eqnarray}
 \mathcal{D} (\hat P)&=&\{\psi,\psi'\in \mathcal{L}^2(-l_{max},\,l_{max}); \psi (-l_{max})=\psi' (l_{max})=0 \}, \\ 
 \mathcal{D} (\hat P^\dag  )&=&\{\psi,\psi'\in \mathcal{L}^2(-l_{max},\,l_{max}) \},
 \end{eqnarray}
	where  $\mathcal{D}(\hat P)$ and $\mathcal{D}(\hat P^\dag)$ respectively  the   dense  domaines of the operator $\hat P$  and $\hat P^\dag$.}\\

{\bf Proof.}  For the proof of this proposition, one can refer to  our  previous reference \cite{1}. \\

Consequently, the scalar product between two states $|\psi\rangle$ and $|\phi\rangle$ and the orthogonality of
position eigenstate become
\begin{eqnarray}
\langle \psi |\phi\rangle&=& \int_{-l_{max}}^{+l_{max}}\frac{dx}{1-\tau  x+\tau^2 x^2} \psi^*(x)\phi(x).\\
\langle x |x'\rangle&=& (1-\tau  x+\tau^2 x^2)\delta(x-x').
\end{eqnarray} 

With the symmetrization of the operator $\hat P$, we have:\\

{\bf Lemma 2.2:} {\it Since the operator $\hat P$ is symmetric,  then all eigenvalues are real and the corresponding  eigenvectors  are orthogonal.}\\

{\bf Proof.} If $\hat P |\psi_i\rangle=\lambda_i |\psi_i\rangle$ \, ($i=1,2$)  with $\ \lambda_i\in \mathbb{C}$, then we  have 
$\langle \hat P^\dag \psi_1|\psi_1\rangle= \langle  \psi_1|\hat P\psi_1\rangle=
\lambda_1^*||\psi_1 ||^2=\lambda_1||\psi_1 ||^2$.
Since  $\lambda_1^*=\lambda_1$, and 
$\langle \hat P^\dag \psi_1|\psi_2\rangle= \langle  \psi_1|\hat P\psi_2\rangle=
(\lambda_1-\lambda_2)\langle \psi_1|\psi_2\rangle=0$ therefore $\langle \psi_1|\psi_2\rangle=0,\quad \mbox{with}\quad  (\lambda_1-\lambda_2)\neq 0$,
finishing the proof.

\subsection{ Deformed  translation  symmetry}
Let us consider for instance the configuration space wave functions in $x$-direction such as  $\psi(x)=\langle x|\psi\rangle $ defined  in the domain $\mathcal{D}_\tau$, the action of the generalized position operator  on this function $\hat P$ yields 
\begin{eqnarray}
\hat P\psi(x)&=&-i\hbar D_x\psi(x),\\
\frac{i}{\hbar}\hat P\psi(x)&=&  D_x\psi(x)\label{1}
\end{eqnarray}
where    $D_x= (1-\tau  x+\tau^2  x^2)\partial_x$ is  the deformed partial derivative  which introduces asymmetrical effect  in the system.\\

{\bf Proposition 2.2:} {\it 
 For the operator $\hat P$, one can associate through the exponential map,  a $\tau$-translation operator $\hat {\mathcal{T}}_\tau$  that induces  an infinitesimal distance $\delta a$ in this direction
\begin{eqnarray}
\hat {\mathcal{T}}_\tau (\delta a)=\exp\left(\frac{i}{\hbar}\delta a\hat P\right),\quad \mbox{with}\quad \delta a \in \mathbb{R}.
\end{eqnarray}}

{\bf Proof:} For infinitesimal translations in space follows from
the Taylor series expansion: 
$\psi(x+\delta a)=\psi(x)+\delta a D_x\psi(x)+\frac{1}{2!} (\delta a)^2 D_x^2\psi(x)+...=\sum_{n=0}^{\infty}\frac{1}{n!}(\delta a D_x)
^n\psi(x)=
\exp\left(\delta a D_x\right) \psi(x)$.
Using the relation (\ref{1}), we have
$	\psi(x+\delta a)=\exp\left(\frac{i}{\hbar}\delta a\hat P\right) \psi(x)$,
finishing the proof.

An alternative way of seeing this result without relying on the wave function representation is by considering
the following action of the exponentiated operator on the  state $|x\rangle$, 
translates this state by the amount of $\delta a$  to the right
\begin{eqnarray}
\exp\left(-\frac{i}{\hbar}\delta a\hat P\right)|x\rangle= |x+\delta a( 1-\tau x+\tau^2 x^2)\rangle.
\end{eqnarray}
From this, it also follows that 
\begin{eqnarray}
\langle x|\exp\left(\frac{i}{\hbar}\delta a\hat P\right)= \langle x+\delta a( 1-\tau x+\tau^2 x^2)|.
\end{eqnarray}
Next, let us characterize the unitary of  this operator.\\

{\bf Lemma 2.2:} {\it  Since the  operator  $\hat P$ is  symmetric $\hat P=\hat P^\dag$, then $\hat {\mathcal{T}}_\tau$    is unitary and  satisfy the  following relations:\\\\
	(i) $ ||\hat {\mathcal{T}}_\tau\psi|| = ||\psi||$,\\
	(ii)  $ \langle  \hat {\mathcal{T}}_\tau\psi|\hat {\mathcal{T}}_\tau\phi \rangle=\langle  \psi|\phi \rangle $,\\
	(iii) $(\hat {\mathcal{T}}_\tau (\delta a))^\dag=\hat {\mathcal{T}} (-\delta a)=\hat ({\mathcal{T}}_\tau(\delta a))^{-1}.$
}\\

{\bf Proof:} (iii)  $\hat P$ is symmetric, then  we have:
 $(\hat {\mathcal{T}}_\tau (\delta a))^\dag= e^{-\frac{i}{\hbar}\delta a\hat P^\dag}=e^{-\frac{i}{\hbar}\delta a\hat P}=\hat {\mathcal{T}} (-\delta a)
 =\left[e^{\frac{i}{\hbar}\delta a\hat P}\right]^{-1}=\hat ({\mathcal{T}}_\tau(\delta a))^{-1}$. Based on this equation, we have:
$(\hat {\mathcal{T}}_\tau (\delta a))^\dag (\hat {\mathcal{T}}_\tau (\delta a))= \hat {\mathcal{T}}_\tau (-\delta a) \hat {\mathcal{T}}_\tau (\delta a)=\exp\left(\frac{i}{\hbar}\delta a\hat P\right)\exp\left(-\frac{i}{\hbar}\delta a\hat P\right)= \mathbb{I}$.
Then (iii)$\implies$  (ii) 
$\langle  \hat {\mathcal{T}}_\tau\psi|\hat {\mathcal{T}}_\tau\phi \rangle=\langle \hat {\mathcal{T}}_\tau^\dag \hat {\mathcal{T}}_\tau\psi|\phi \rangle=\langle  \psi|\phi \rangle.$
Finally, (ii) $\implies$ (i) as follows:
$||\hat {\mathcal{T}_\tau} \psi||^2=\langle  \hat {\mathcal{T}}_\tau\psi|\hat {\mathcal{T}}_\tau\psi \rangle =\langle  \psi|\psi \rangle=||\psi||^2,$
finishing the proof.

Let us consider $\hat H$,  the operator Hamiltonian of a system defined within this  space  such as 
\begin{eqnarray}\label{ham}
\hat H (\hat P,\hat X)=\frac{\hat P^2}{2m} +V(\hat X),
\end{eqnarray}
where $V(x)$ is the time-independent  potential energy of the system. For  the Hamiltonians that describe lattice systems, one can set the following theorem:\\

{\bf Theorem 2.2:} {\it For periodical  deformed  potential energy  such  as 
	\begin{eqnarray}
		V(\hat x+\delta a(1-\tau \hat x+\tau^2 \hat x^2))=V(\hat x)
	\end{eqnarray}
then, 
\begin{eqnarray}
[\hat H,\hat {\mathcal{T}}_\tau]=0.
\end{eqnarray}}

{\bf Proof:} Considering that:
$ \hat X \hat {\mathcal{T}}_\tau(\delta a)|x\rangle= (x+\delta a(1-\tau  x+\tau^2  x^2))|x+\delta a(1-\tau  x+\tau^2  x^2)\rangle$
and 
$\hat {\mathcal{T}}_\tau(\delta a)(\hat X+\delta a(1-\tau \hat X+\tau^2 \hat X^2)) |x\rangle=( x+\delta a(1-\tau  x+\tau^2  x^2))|x+\delta a(1-\tau  x+\tau^2  x^2)\rangle.$
From these equations, we  obtain the following  relation:
$\hat {\mathcal{T}}_\tau(\delta a)(\hat X+\delta a(1-\tau \hat X+\tau^2 \hat X^2))=\hat X \hat {\mathcal{T}}_\tau(\delta a).$
From this equation, on can progressively deduce that
$\hat {\mathcal{T}}_\tau(\delta a)V(\hat X+\delta a(1-\tau \hat X+\tau^2 \hat X^2))=V(\hat X) \hat {\mathcal{T}}_\tau(\delta a)$.
So, if  the potential energy $V(\hat X)$  is periodic, therefore  it  is invariant under this deformed translation such as:
$V(\hat X+\delta a(1-\tau \hat X+\tau^2 \hat X^2))=V(\hat X)=
V(\hat x+\delta a(1-\tau \hat x+\tau^2 \hat x^2))=V(\hat x)$.
Since $[\hat {\mathcal{T}}_\tau ,\hat P]=0$, then we get the following commutation relation: 
$[\hat H,\hat {\mathcal{T}}_\tau]=0$,
finishing the proof.

The time-dependent deformed Schrödinger equation is given by
\begin{eqnarray}\label{H}
\hat H  \psi (x,t)=-\frac{\hbar^2}{2m} D_x^2 \psi (x,t)+V(x)\psi (x,t)= i\hbar \partial_t \psi (x,t).
\end{eqnarray}
 If the  wave function $\psi(x,t)$ is normalized, it
is possible to define a probability density $\rho(x,t)=|\psi(x,t)|^2$.
Using Eq. (\ref{H}) it is straightforward to derive a modified
continuity equation
\begin{eqnarray}
\frac{\partial \rho(x,t)}{\partial t}+D_x J_\tau(x,t)=0,
\end{eqnarray}
where the current density is given by
\begin{eqnarray}
J_\tau(x,t)=\frac{\hbar(1-\tau x+\tau^2 x^2)}{2im}\left(\psi^*\frac{d\psi}{dx}-\psi\frac{d\psi^*}{dx}\right).
\end{eqnarray}
With these equations at hand, we are now in a position to determine the eigensystems of a particle in an infinite square well potential at the Planck scale and to deduce later from this spectrum its thermodynamic properties.

\section{ Spectrum of  particle in an infinite square-well potential}\label{sec3}
Let us consider  the Hamiltonian of  the above  quantum system  confined in an infinite square well potential at the Planck scale, defined as
\begin{eqnarray}
V(x)=
\begin{cases}
0,\quad   -l_{max}<x<l_{max},  \\
\infty,\quad \mbox{ otherwise}.
\end{cases}
\end{eqnarray}
For standing waves in a null potential, the wave function $\psi(x)$
satisfying equation (\ref{H}) obeys
\begin{eqnarray}\label{eq1}
-\frac{\hbar^2}{2m} D_x^2 \psi (x)=E\psi(x),
\end{eqnarray}
or 
\begin{eqnarray}\label{eq1}
-\frac{\hbar^2}{2m}\left[(1-\tau  x+\tau^2 x^2)^2\frac{d^2}{dx^2}- \tau(1-2\tau x)(1-\tau  x+\tau^2 x^2)\frac{d}{dx}\right] \psi (x)=E \psi (x).
\end{eqnarray}
The solution of the equation (\ref{eq1}) in this  infinite  square-well potential is  given by
\begin{eqnarray}\label{sol1}
\psi_k(x)= A\exp\left(i\frac{2k}{\tau\hbar\sqrt{3}}
\left[\arctan\left(\frac{2\tau x-1}{\sqrt{3}}\right)+\arctan\left(\frac{3}{\sqrt{3}}\right)\right]\right),
\end{eqnarray}
where $k=\frac{\sqrt{2mE}}{\hbar} $ and $A$ is a constant. Then by normalization, $\langle \psi_k|\psi_k\rangle=1$, we have 
\begin{eqnarray}
1&=&\int_{-l_{max}}^{+l_{max}}\frac{dx}{1-\tau x+\tau^2 x^2} \psi^* (x) \psi (x)\cr
&=& A^2 \int_{-l_{max}}^{+l_{max}}\frac{dx}{1-\tau  x+\tau^2 x^2}.
\end{eqnarray}
so, we find
\begin{eqnarray}\label{nzeta}
A=\sqrt{\frac{\tau \sqrt{3}}{\pi}}.
\end{eqnarray}
Substituting this equation (\ref {nzeta}) into the equation (\ref{sol1}), we have 
\begin{eqnarray}\label{resp}
\psi_k (x) &=&  \sqrt{\frac{\tau \sqrt{3}}{\pi}}\exp\left(i\frac{2k}{\tau\hbar \sqrt{3}}\left[\arctan\left(\frac{2\tau x-1}{\sqrt{3}}\right)
+\arctan\left(\frac{3}{\sqrt{3}}\right)\right]\right).
\end{eqnarray}

Based on the reference  \cite{25''b}, the scalar product of the formal  eigenstates  is given by
\begin{eqnarray}
\langle \psi_{k'}|\psi_k\rangle&=&\frac{\tau \sqrt{3}}{\pi(k-k')}e^{i\frac{2(k-k')}{\tau \hbar \sqrt{3}}\arctan\left(\frac{3}{\sqrt{3}}\right)}\cr&&\times\left[e^{i\left(\frac{2(k-k')}{\tau \hbar \sqrt{3}}\arctan\left(\frac{2\tau l_{max}-1}{\sqrt{3}}\right)-\frac{\pi}{2}\right)}-e^{-i\left(\frac{2(k-k')}{\tau \hbar \sqrt{3}}\arctan\left(\frac{2\tau l_{max}+1}{\sqrt{3}}\right)+\frac{\pi}{2}\right)}\right].
\end{eqnarray}

\begin{figure}[htbp]
	\resizebox{1.2\textwidth}{!}{
		\includegraphics{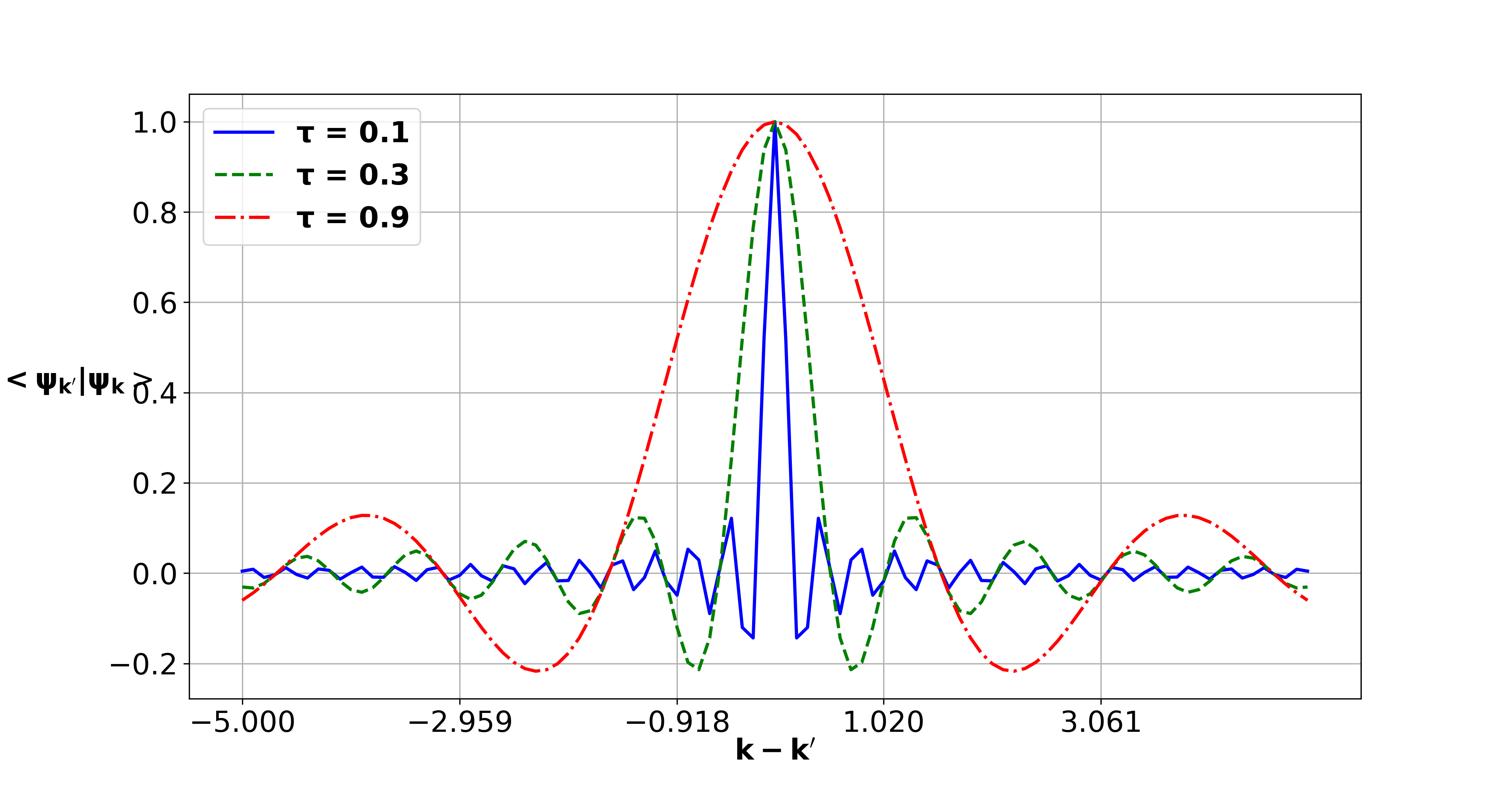}
	}
	\caption{\it \small  Variation of $\langle \psi_{k'}|\psi_k\rangle$ versus k-k' with $\hbar=1$.
	}
	\label{fig}       
\end{figure}

This relation  shows that, the normalized  eigenstates (\ref {resp}) are  no longer
orthogonal. However,  if one tends $(k-k')\rightarrow \infty$, these states  become orthogonal 
\begin{eqnarray}
\lim_{(k-k')\rightarrow \infty} \langle \psi_{k'}|\psi_{k}\rangle=0.
\end{eqnarray}
These  properties show that,  the states $|\psi_k\rangle$ are   essentially Gaussians centered at  $(k-k')\rightarrow 0$ (see Figure \ref{fig}). They can be assimilated to the coherent states \cite{25'''}  which are known as states that mediate a smooth transition between the quantum and classical worlds. This transition is manifested by the saturation of the Heisenberg uncertainty principle $\Delta_z q\Delta_z p=\frac{\hbar}{2}$. 
In comparison with    coherent
states, the states $|\psi_k\rangle$ strongly saturate the GUP ($\Delta_\psi X\Delta_\psi P=\hbar$) at the Planck scale and could be used to describe the transition states between the quantum world and unknown world for which the physical descriptions are out of reach.

We impose  that the wave function satisfies the  Dirichlet condition i.e it vanishes at the boundaries    $\psi_k(-l_{max})=0=\psi_k(+l_{max})$.
Thus, using especially the boundary condition $\psi(-l_{max})=0$,  the 
above wave functions (\ref{resp}) becomes
\begin{eqnarray}
\psi_k (x) =  \sqrt{\frac{\tau\sqrt{3}}{2}}\sin\left(\frac{2k}{\tau \sqrt{3}}\left[\arctan\left(\frac{2\tau x-1}{\sqrt{3}}\right)
+\frac{\pi}{3}\right]\right).
\end{eqnarray}
The quantization follows from the boundary condition $\psi(l_{max})=0$ and leads to the equation
\begin{eqnarray}
\frac{2k_n}{\tau \sqrt{3}}\left[\arctan\left(\frac{2\tau l_{max}-1}{\sqrt{3}}\right)
+\frac{\pi}{3}\right]=n\pi\quad \mbox{with}\quad n\in \mathbb{N}^*.
\end{eqnarray}
Then, the  energy spectrum of the particle  is written as
\begin{eqnarray}
E_n &=& \frac{3\pi^2\hbar^2n^2}{8                    m }\tau^2 \left[\arctan\left(\frac{2\tau l_{max}-1}{\sqrt{3}}\right)
+\frac{\pi}{3}\right]^{-2}\cr
&=& 3\frac{\hbar^2\tau^2 n^2}{2m}.\label{sp1}
\end{eqnarray}
In term of the maximal  length of quantum gravity, we have 
\begin{eqnarray}
E_n =\frac{3}{\pi^2}\frac{\hbar^2 \pi^2 n^2}{2m l_{max}^2}=0.3 \frac{\hbar^2 \pi^2 n^2}{2m l_{max}^2}.
\end{eqnarray}
From this result, if we assume that  the maximal length  $l_{max}$ is the order of   the ordinary length $L$ of square-well potential in the  basic  quantum mechanics i.e $l_{max}=L$, the spectrum of this system is expressed as follow
\begin{eqnarray}
E_n =0.3\varepsilon_n\implies E_n < \varepsilon_n,
\end{eqnarray} 
where $  \varepsilon_n=\frac{\hbar^2  \pi^2 n^2}{2m L^2}$ is the spectrum of a free  particle in an infinite  square well potential of the basic quantum mechanics with the fundamental energy 
$  \varepsilon_1=\frac{\hbar^2\pi^2 }{2m L^2}$. This result shows that the strong  graviton measurement induces weak transition energies.  This
indicates that, the quantum gravity induces a more pronounced contraction of
energy levels which, consequently implies the decrease of energy band structures \cite{{25a'}}.

Hereafter, with the spectrum (\ref{sp1}) we are able  to study in detail the statistical properties of this system  at the Planck scale.

\section{ Thermodynamic descriptions }\label{sec4}

We consider canonical and microcanonical ensembles of ideal gas composed of the above systems, each of them consists of N identical noninteracting particles enclosed by the adiabatic wall with constant volume V. We suppose that the systems are in thermal equilibrium with their surroundings at temperature T. 

Since the above system is strongly influenced by the maximal presence of graviton, therefore the thermodynamic quantities such as the internal energy E, the Helmholtz free energy A, the entropy S, and the chemical potential M  could be also strongly modified or corrected. Furthermore, since the partition function is the key entity in the description of a canonical ensemble,  we use this approach to compute the modified thermodynamic properties E, A, S, and M.

\subsection{ The partition function method}

   The partition function of a single molecule of a system is given by:
\begin{eqnarray}\label{part1}
Z=\sum_{n=0}^{+\infty}\exp\left(-\beta E_n\right),
\end{eqnarray}
where $\beta =\frac{1}{k_B T}$, $k_B$ is the Boltzmann constant and $T$ represents the thermodynamic temperature. 
Inserting  Eq.(\ref{sp1}) in Eq.(\ref{part1}) yields:
\begin{eqnarray}\label{part2}
Z= \sum_{n=0}^{+\infty}\exp\left(-\beta\frac{3\hbar^2\tau^2}{2m }n^2\right).
\end{eqnarray}
In view of the largeness of the number of states of the particles and the largeness of the volume of
the well to which the particles are confined, one may regard the number of states in each coordinate
direction as a continuous variable. In this case the Eq.(\ref{part2}) becomes:
\begin{eqnarray}
Z=\int_{0}^{\infty}dx \exp\left(-\beta\frac{3\hbar^2\tau^2}{2m }x^2\right).
\end{eqnarray}
The integration yields
\begin{eqnarray} \label{part3}
Z&=&   \frac{z_0}{\tau },  
\end{eqnarray}
where $z_0=\frac{1}{\hbar}\sqrt{\frac{2\pi m}{3\beta}}$  is the ordinary partition function for a single particle.
The partition function of the N distinguishable particle systems then can be defined as
\begin{eqnarray}\label{part4}
Z_N=\frac{(Z)^N}{N!}
\end{eqnarray}
Inserting Eq.(\ref{part3}) in Eq.(\ref{part4}), the partition function becomes:
\begin{eqnarray}
Z_N= \frac{z_0^N}{N!\tau^N } =  \frac{z}{\tau^N },
\end{eqnarray}
where    $z=\frac{z_0^N}{N! }$ represents the ordinary partition function and $Z_N$ is the $\tau$-modified partition function of the  system. 

From the partition function $Z_N$, the modified  internal energy is deduced as follows 
\begin{eqnarray}
E=-\left(\frac{\partial \ln Z_N}{\partial \beta}\right)_{N}=\frac{1}{2}Nk_BT.
\end{eqnarray}
This equation shows that the internal energy is not influenced by the maximal length of the graviton. This observation comes to confirm the recently obtained result with Perivolaropoulos's algebra \cite{26}. In contrast with the obtained results in the presence of a minimal length, \cite{20,21,22,23,24,25}, the ordinary internal energy is shifted at the first order of deformed parameter and led to its decreasing. Thus,
the  specific heat  $C$  remains also unchanged and is given by
\begin{eqnarray}
C=\left(\frac{\partial E}{\partial  T}\right)_{N}= \frac{1}{2}Nk_B.
\end{eqnarray}

Concerning the  modified Helmholtz free energy A, it has the expression
\begin{eqnarray}\label{hel}
A&=&-k_B T \ln Z_N\cr
&=& a+ Nk_BT\ln\tau,
\end{eqnarray}
where $a=- k_B T \ln z$ is the ordinary Helmholtz free energy. This shows that 
the ordinary Helmholtz free energy is corrected by a logarithm of the deformed parameter such as   $\Delta A=  Nk_BT\ln\tau$. 

Another important quantity that we can deduce  from the equation (\ref{hel} ) is the modified entropy defined as
\begin{eqnarray}
S&=&-\left(\frac{\partial A}{\partial T}\right)_{N}\cr
&=&  s-  Nk_B\ln\tau.\label{entr1}
\end{eqnarray}
where the first term $s=-\left(\frac{\partial a}{\partial T}\right)_{N,T}$ is the ordinary entropy and the second term represents
the correction to the  ordinary entropy  $\Delta S=- Nk_BT\ln\tau $. Note that this logarithmic correction to the ordinary entropy is known in various approaches of the extremal limit of a black hole geometries \cite{16,17,18,19}. According to several works in this frame, the entropy logarithmic correction is due to large quantum fluctuations at this scale. This situation is consistent with our context (\ref{alg1}) since the quantum gravity measurement is important at this scale.

Finally, the generalized chemical potential, M is given by
\begin{eqnarray}
M&=& \left(\frac{\partial A}{\partial N}\right)_{T}\cr
&=& \mu + k_BT\ln\tau,
\end{eqnarray}
where $  \mu=\left(\frac{\partial a}{\partial N}\right)_{T}$ stands for the chemical potential in the undeformed case and  $ \Delta M =  k_BT\ln\tau$ is the induced  logarithmic chemical correction.  

In summary of the above logarithmic corrections due to  the  strong gravity at this scale, we have: 
\begin{eqnarray}
\Delta A=   Nk_BT\ln\tau, \quad  \Delta S=  -Nk_B\ln\tau  \quad \mbox{and} \quad \Delta M =  k_BT\ln\tau.
\end{eqnarray}
Note that  at the  extremal limit of the space ( $\tau \rightarrow 0$), 
the above  thermodynamic  quantities divergent  such as  
\begin{eqnarray}
\lim_{\tau\rightarrow 0 } \Delta A&=&\infty\implies  \lim_{\tau\rightarrow 0 } A=\infty.\\
\lim_{\tau\rightarrow 0 } \Delta S&=&\infty \implies  \lim_{\tau\rightarrow 0 } S=\infty.\\
\lim_{\tau\rightarrow 0 } \Delta M&=&\infty \implies  \lim_{\tau\rightarrow 0 } M=\infty.
\end{eqnarray}
Comparing these results with the analogous obtained in the minimal length scenarios \cite{20,21,22,23,24,25} show that,  the effect of the maximal length in this framework and that of the minimal length are fundamentally different. In fact, the minimal length induces a weak graviton localization while the maximal length in this context induces a strong graviton localization. The weak graviton measurement shifts quadratically the thermodynamic quantities and at the extremal limit of deformed parameter $\beta_{GUP}$  $(\beta_{GUP}\rightarrow 0)$, one recovers the ordinary quantities whereas, in this scenario, the graviton measurement induces logarithmic corrections.
\subsection{ The density of states method}
Another important thermodynamic characteristic is the density of states. The density of states  $G_N(E)$ around the energy value $E$ is obtained  as the
the inverse of Laplace transform of the   partition function $Z_N$ \cite{27}
\begin{eqnarray}\label{den1}
G_N(E)=\frac{1}{2\pi i}\int_{\beta'-i\infty}^{\beta'+i\infty}e^{\beta E} Z(\beta)d\beta,\quad \beta'>0.
\end{eqnarray}
Inserting the partition function (\ref{part3}) in (\ref{den1})  results in
\begin{eqnarray}
G_N(E)=\frac{1}{ N!(\hbar\tau)^N}\left(\frac{2\pi m}{3}\right)^{\frac{N}{2}}\frac{1}{2\pi i}\int_{\beta'-i\infty}^{\beta'+i\infty} \frac{e^{\beta E}}{\beta^{\frac{N}{2}}}d\beta.
\end{eqnarray}
This integral can be calculated by using the formula:
\begin{eqnarray}
 I=\begin{cases}
 \frac{1}{2\pi i}\int_{\beta'-i\infty}^{\beta'+i\infty} \frac{e^{\beta x}}{\beta^{n+1}}d\beta=\frac{x^n}{n!},\quad  \mbox {for} \quad  x> 0,\\\\
 0  \quad  \mbox {for} \quad  x< 0.
 \end{cases}
\end{eqnarray}
Based on this integral, one gets

\begin{eqnarray}\label{den2}
G_N(E)=\begin{cases}
\frac{1}{ N!(\hbar\tau)^N}\left(\frac{2\pi m}{3}\right)^{\frac{N}{2}}\frac{E^{\frac{N}{2}-1}}{\left({\frac{N}{2}-1}\right)!},\quad  \mbox {for} \quad  E> 0,\\\\
0  \quad  \mbox {for} \quad  E< 0. 
\end{cases}
\end{eqnarray}
Let us consider $g_N(E)= \frac{1}{ N!}\left(\frac{2\pi m}{3\hbar^2}\right)^{\frac{N}{2}}\frac{E^{\frac{N}{2}-1}}{\left({\frac{N}{2}-1}\right)!}$, the ordinary density of states, the generalized density of states (\ref{den2}) can be written as
\begin{eqnarray}
G_N(E)= \frac{g_N(E)}{\tau^N},\quad \mbox{for}\quad E>0.
\end{eqnarray}
This result is not only different from that of the minimal length regime \cite{28}, but is also different to that of obtaining with Perivolaropoulos's maximal length \cite{26}.  

The  generalized number of microstates accessible to the system is given by
\begin{eqnarray}\label{mi1}
\Omega (E)= \frac{\omega(E)}{\tau^N}, \quad \mbox{for}\quad E>0,
\end{eqnarray}
where $ \omega(E)= \frac{1}{ N!}\left(\frac{2\pi m}{3\hbar^2}\right)^{\frac{N}{2}} \frac{E^{\frac{N}{2}-1}}{\left({\frac{N}{2}-1}\right)!}\delta E $ is  the ordinary number of microstates accessible to the system with energy lying between $E$ and $E+\delta E$ .

The microcanonical ensemble (\ref{mi1}) defined through the parameters N, and E allows the  study of thermodynamic quantities by the use of the deformed entropy
\begin{eqnarray}
S=k_B\ln \Omega=s-Nk_B\ln\tau.
\end{eqnarray}
where $s$ is the ordinary microcanonical entropy. This expression is exactly the same given by Eq.(\ref{entr1}). It is important to notice that under the effect of the strong graviton,  the deformed entropy induces a logarithm correction that decreases the number of microstates accessible to the system.
From this  entropy, one gets the same expressions
established for the rest of the thermodynamic properties in the previous subsection.
The internal energy is given by
\begin{eqnarray}
	dE=TdS\implies E=\frac{1}{2}Nk_BT.
\end{eqnarray}
 Concerning,  the corrected chemical potential, one gets
 \begin{eqnarray}
 M&=& -T\left( \frac{\partial S}{\partial N}\right)_E\cr
    &=&\mu + T k_B\ln\tau,
 \end{eqnarray}
 where $\mu= -T\left( \frac{\partial s}{\partial N}\right)_E$ is  the undeformed chemical potential.
 
 Finally, the  corrected Helmholtz free energy is related to the internal energy and the  entropy via the following relation relationship 
 \begin{eqnarray}
 	A&=&E-TS\cr
 	&=& a+ Nk_BT\ln\tau
 \end{eqnarray}
 where $ a= E-Ts  $  is the undeformed Helmholtz free energy

As expected, similar to the ordinary case, both canonical and microcanonical formalisms yield identical results in the presence of a maximum length.

\section{Conclusion}\label{sec5}

In this paper, we studied the effects of maximal measurement of quantum gravity on the thermodynamics of canonical and microcanonical ensembles of N noninteracting particles at the Planck scale. These showed that the canonical ensemble based on the use of partition function and the microcanonical ensemble based on the density of states lead to the same result. Before computing the thermodynamic quantities, we analytically determined the spectrum of a particle in a 1D infinite square well potential in the framework of our recent position-dependent deformed algebra  \cite{1}. We showed that the energy spectrum of this system is weakly proportional to the ordinary one of quantum mechanics free of gravity theory. This characterization of quantum gravity is in perfect
 analogy with the one of General Relativity (GR) point of view.

The interesting physical result we found in the present paper is that,  at the Planck scale with strong quantum gravity localization, the ordinary thermodynamic quantities as the Helmholtz free energy $a$, the entropy $s$, and the chemical potential $\mu$   shift logarithm corrections of the deformed parameter $\tau$, except the internal energy which is invariant under the strong influence of graviton. The logarithm corrections induced by these quantities are consistent with the ones obtained at the extremal limit of black holes \cite{16,17,18,19} due to the high effectiveness of quantum fluctuation \cite{29}. To confirm these results, we compared our results with those of the minimal length formalism, available in the literature \cite{20,21,22,23,24,25}. On this basis, it is shown that both scenarios are fundamentally different.  In fact, the minimal length formalism shifts quadratically the thermodynamic quantities at the order of the deformed parameter $\beta_{GUP}$.
So,  at the extremal limit i.e  $\beta_{GUP}\rightarrow 0$ one recovers the ordinary quantities while in this framework by tending $\tau \rightarrow 0$, the modified thermodynamic quantities divergent because of the logarithm corrections. With respect to these remarks, we could state that the minimal length formalism \cite{20,21,22,23,24,25} induced weak gravitons localization while the emergence of maximal length in this algebra induced strong graviton at the Planck scale.  However, referring to the results of  Bensalem and  Bouaziz \cite{26} where the authors computed the thermodynamic quantities of the same model in the framework of maximal length introduced by Perivolaropoulos \cite{15}. They were coming to the conclusion that the GUPs with the maximal length and minimal length lead to the same effects on the same thermodynamic quantities. In view of this conclusion, one could postulate that the GUP with maximal length induces weak graviton localization, which at first,  seems to be counterintuitive with our proposal. Let us clarify that, the difference in the meaning of both studies is related to the forms of the introduced deformed algebras.   Perivolaropoulos's algebra \cite{15} used to compute these thermodynamic quantities \cite{26} is corrected at the first order of the deformed parameter $\alpha$ while my version (\ref{alg1}) is quadratically corrected until the second order of the deformed parameter $\tau$. So, the effects of quantum gravity increase with the order of the deformed parameter.

\section*{Acknowledgments}
 The author acknowledges support from AIMS-Ghana
under the Postdoctoral fellow/teaching assistance (Tutor) grant

\end{document}